\documentstyle{article}
\begin{document}

\begin{center}
{\Large \bf Polynomial Gauge Invariants of a Bosonic String} \\

\vspace{4mm}

Vasiliy Dolgushev\footnote{e-mail vald@phys.tsu.ru}
\\
Physics Department, Tomsk State University,
Tomsk 634050 Lenin ave 36 \\
RUSSIA \\
\end{center}

\begin{abstract}
An open bosonic string is considered with the aim to
construct a general gauge invariant, being a polynomial of
Fubini-Veneziano (FV) fields.  The FV fields are transformed
 as 1-forms on $S^1$\,, that allows to formulate the problem in geometric
terms. We introduce a most general anzats for these invariants and
explicitly resolve the invariance conditions in the framework of the anzats.
The invariants are interpreted as integrals of n-form over a gauge invariant
domains in an n-dimensional torus, where the invariance of these domains is
considered with respect to the action of the diagonal of the group
$\times (Diff~S^1)^n$.

We also discuss a possibility to get a complete set of gauge invariants
which allow an actual dependence on the string zero modes.  We find
that the complete set can't be restricted by polynomial invariants only.

The classical polynomial invariants, being directly defined in the
string Fock space, turn out to break the structure of the respective BRST
cohomology even in the critical dimension. We discuss a possibility to
restore the BRST invariance of the corresponding operator algebra by a
non-trivial quantum deformation of the original invariants.
\end{abstract}

\section*{Introduction}
The aim of the work is to find a complete set of gauge invariants of a
bosonic string.

A classical gauge invariant is understood as a parametrization
independent object that is a physical observable. And a quantum gauge
invariant is an operator which is well-defined in the respective
BRST-cohomology. It is an operator which represents a physical
observable.

A complete set of classical gauge invariants is defined as the
set in terms of which an arbitrary physical observable can be
expressed. And a complete set of quantum gauge invariants
is the set whose enveloping algebra includes all invariant
operators. It will be shown how to find all classical invariants at
least in the class of polynomials. The question of the
quantization of these invariants will be also discussed.

The bosonic string is a well studied model. It allows to apply various
methods of quantization, it's spectrum can be obtained in different
ways. However a structure of the reduced phase space of the model is
rather complicated and is not recognized well.
It is the set of gauge invariants that can be applied for investigating
the structure.

It could be useful for constructing string interaction, for a string
field theory. There is also another question less obvious and less
well known, namely to understand how the phase space of the string
stratifies into phase spaces of the elementary particles constituting
it's spectrum. The information about the invariants seems to be rather
useful for elaborating the last question.

It is commonly known that the complete set of quantum gauge invariant
can be represented by the set of vertex operators \cite{Green}.
However the vertex operators have no classical limit.
And actually we are looking for the another set of invariants which
do have a certain classical limit.

For simplicity we'll restrict ourselves to the case of the open bosonic
string.

\section*{Classical gauge invariants}
The complete set of the phase space variables of the open bosonic
string consists of the Fubini-Veneziano(FV) fields and the string zero
mode
\begin{equation}
V^{\mu}=V^{\mu}(\sigma)~:~[0\,,2\pi]\rightarrow {\bf R}^{1\,,D-1}
\qquad V^{\mu}(0)=V^{\mu}(2\pi)
\qquad  q^{\mu}\in {\bf R}^{1\,,D-1}
\end{equation}
They are subject to the first class constraints
\begin{equation}
L(\sigma)=\frac{1}{4}V^{\mu}(\sigma)V_{\mu}(\sigma)
\end{equation}

First of all let us pose the question whether there are polynomial
gauge invariants which depend on the FV fields only.
The positive answer to the question can be found in the literature,
namely an infinite set of such invariants was proposed in the works of
Pohlmeyer and Rehren \cite{1,2,3}.
\begin{equation}
I_n^{\mu_1\mu_2\ldots\mu_n}=
\int_0^{2\pi}V^{\mu_1}(\sigma_1) d\sigma_1
\int_{\sigma_1}^{\sigma_1+2\pi}
d\sigma_2V^{\mu_2}(\sigma_2)
\int_{\sigma_1}^{\sigma_2}
V^{\mu_3}(\sigma_3)
\ldots \int_{\sigma_1}^{\sigma_{n-1}}d\sigma_n
V^{\mu_n}(\sigma_n)
\label{Po}
\end{equation}
In the paper \cite{3} it was proved that these polynomials (\ref{Po})
exhaust all gauge invariants which depend on the FV fields only.
If we do need to obtain a complete set of classical gauge invariants we
should involve an actual dependence on the string zero mode
$q^{\mu}$\,.

The most general polynomial expression for a classical gauge invariant
is as follows
\begin{equation}
I=\sum
C_{\mu_1\mu_2\ldots \mu_n}q^{\mu_1}q^{\mu_2}\ldots q^{\mu_n} +\\
\sum C_{{\nu_1}\mu_1\mu_2\ldots \mu_{n-1}}^{m_1}\alpha_{m_1}^{\nu_1}
q^{\mu_1}q^{\mu_2}
\ldots q^{\mu_{n-1}}+\ldots
+\sum C_{\nu_1\nu_2\ldots\nu_n}^{m_1 m_2 \ldots m_n}
\alpha_{m_1}^{\nu_1} \alpha_{m_2}^{\nu_2}\ldots \alpha_{m_n}^{\nu_n}
\label{Inv}
\end{equation}
where
$$\alpha^{\mu}_{n}=\frac{1}{2\sqrt{\pi}}\int_0^{2\pi}
V^{\mu}(\sigma)e^{-in\sigma}d\sigma$$
As we see all the terms in the expression (\ref{Inv}) are of the same
order in the phase space variables. One can take the anzats in such a
form simply because the gauge
transformations are homogeneous in the phase space variables:
\begin{equation}
\delta_{\varepsilon}
V^{\mu}
=(\varepsilon(\sigma)V^{\mu}(\sigma))'
\qquad
\delta_{\varepsilon}
q^{\mu}=\int_0^{2\pi}d\sigma\varepsilon(\sigma)
V^{\mu}(\sigma)
\qquad
\delta I=\{\,L[\varepsilon] \,,I \}
\label{gauge}
\end{equation}
If one requires the polynomial (\ref{Inv}) to be the gauge invariant
the respective structure coefficients are subject to following
conditions:
$$ C_{\mu_1\mu_2\ldots\mu_n}=0\,,$$
$$  C_{\nu_1\nu_2\ldots \nu_l
\mu_1\ldots\mu_{n-l}}^{n_1 n_2\ldots n_l}=0
$$
if
$$ n_1\neq 0\,,n_2\neq 0\,,\ldots\,,n_l\neq 0\,,$$
\begin{equation}
C_{\nu_1\nu_2\ldots \nu_l
(\nu_{l+1}\ldots \nu_s\mu_1)\ldots\mu_{n-s}}^{n_1 n_2\ldots n_l\, 0~
\ldots ~0}=0\,.
\label{aha}
\end{equation}
One of the examples of such invariants which depends on the whole set of
phase space variables is the momentum tensor of the string:
\begin{equation}
{\cal M}^{\mu\nu}= q^{\mu}\alpha_0^{\nu}-q^{\nu}\alpha_0^{\mu}+
\sum_{n\neq 0}\frac{i}{n}\alpha_n^{\mu}\alpha_{-n}^{\nu}
\label{Poinc}
\end{equation}

Using the relations (\ref{aha})
one proves that an abitrary polynomial gauge invariant can be
expressed, modulo constraints, in terms of the momentum
tensor (\ref{Poinc}) and the polynomials (\ref{Po}).
It turns out that the
proposed polynomial invariants form only a subalgebra of
the algebra of physical observables. Actually they do not exhaust the
complete set of string gauge invariants because there are physicaly
different points on the constraint surface of the string, that cannot
be distinguished with the help of these polynomials. The last means
that the complete set of string gauge invariant must include
observables  which are not polynomial in the phase space variables.
Unfortunately no one of such invariant is known yet.

\section*{Quantization problem}
Let us discuss a quantization of the polynomial invariants.
As we know the momentum tensor (\ref{Poinc}) of the bosonic string can
be quantized with out any problems. The same situation takes place with
the polynomials (\ref{Po}) while $n<4$\,. As to the invariants
(\ref{Po}) with $n$ being more or equaled 4 the situation drastically
changes. Namely the invariants being directly defined in the Fock space
of the string do not commute with the Virasoro generators because of
the quantum corrections. It means that the respective operators are not
defined in the space of the physical states.
The given situation relates to the common quantization problem of
the systems with constraints.  It would be rather strange if
quantum corrections did not destroy some key relations of a classical
theory.
In some cases it leads us to the true values of the critical
parameters, in other ones it means that it is not possible to construct
a consistent quantum theory. There is however the third case when we
simply can say that some relations do not have a consistent
quantum interpretation, but the quantum theory does exist.

We think that the problem we face with can be solved.
Firstly let's note that the classical invariants are defined
ambiguously off the constraint surface. Namely one can add to the
previous polynomial an expression, which vanishes on constraints.
The terms, which vanish classically, may contribute to the
quantum commutator between the invariant and the BRST charge.
\begin{equation}
\Omega=\sum_n L_n C_{-n}+\sum_{n\,m}m P_nC_mC_{-n-m}\,,
\end{equation}
where $C_{-n}$ and $P_n$ are canonical
ghosts and
\begin{equation}
L_n=\frac{1}{2}\sum_k \alpha_k^{\mu}\alpha_{n-k}^{\nu}\eta_{\mu\nu}
\label{op}
\end{equation}
are the Virasoro generators.

And it is the arbitrariness that can be used for constructing genuine
quantum BRST invariants polynomial in string operators and ghosts.

Let us summarize the things to be done.\\
i)It is necessary to add to the naive invariant the most general
expression which vanishes on constraints
\begin{equation}
\begin{array}{c}
I=I[V] \qquad
[\,I, L_n\, ]=\sum_m W_{nm}L_m \\
\hspace*{-0.6cm}\rule[1.5mm]{0.4pt}{1.5mm}\rule[1.5mm]{.4cm}{.4pt}
\rule[1.5mm]{.4pt}{1.5mm}
\end{array}
\end{equation}
ii) to construct a quantum operator with ghosts using the BFV method
$$
\tilde I=I+C_{-m}P_n  W_{nm}+\ldots\,,
$$
iii) to evaluate the commutator between the constructed operator and
the BRST charge
\begin{equation}
\begin{array}{l}
[\tilde
I,\Omega]=\sum_{n}[\,I,L_{n}\,]C_{-n}+
\sum_{n\,k\,l}m
[W_{kl}C_{-l}P_{k}\,,L_{n}C_{-n}]+
\sum_{n\,m\,k\,l}m
W_{kl}[C_{-l}P_{k}\,,P_nC_mC_{-n-m}]+ \ldots \\
\hspace*{2.1cm}\rule[4mm]{0.4pt}{1.5mm}\rule[4mm]{.4cm}{.4pt}
\rule[4mm]{.4pt}{1.5mm}
\hspace*{-0.5cm}\rule[3mm]{0.4pt}{2.5mm}\rule[3mm]{.4cm}{.4pt}
\rule[3mm]{.4pt}{2.5mm}
~
\hspace*{4cm}\rule[4mm]{0.4pt}{1.5mm}\rule[4mm]{.8cm}{.4pt}
\rule[4mm]{.4pt}{1.5mm}
\hspace*{-0.5cm}\rule[3mm]{0.4pt}{2.5mm}\rule[3mm]{.8cm}{.4pt}
\rule[3mm]{.4pt}{2.5mm}
~
\hspace*{4cm}\rule[4mm]{0.4pt}{1.5mm}\rule[4mm]{.8cm}{.4pt}
\rule[4mm]{.4pt}{1.5mm}
\hspace*{-1cm}\rule[3mm]{0.4pt}{2.5mm}\rule[3mm]{.8cm}{.4pt}
\rule[3mm]{.4pt}{2.5mm}
~
\end{array}
\end{equation}

where $\ldots$ means terms with higher structure functions.

While doing the last it is necessary to account only one-loop
contributions because higher corrections are simply vanishing.
At last we can obtain the equation for the additional terms.
These equation could be solvable because the ghost terms give quantum
corrections of the same order as those arisen in the anomaly.

\section*{Conclusion}
Thus we have the set of physical observables which exhaust all
polynomial invariants. We have proved that the set of the polynomial
invariants is not complete.
Also we pose the question whether it is
possible to realize the BRST cohomology of the string operator algebra
in terms of the operators polynomial in string modes and ghosts.
These invariants unlike the vertex operators
can clarify the connection between the reduced phase space of the
bosonic string and the BRST cohomologies of the corresponding quantum
theory \cite{Garl} .



\end{document}